\documentclass[pra,aps,superscriptaddress,twocolumn]{revtex4-2}

\usepackage{amsmath}
\usepackage{amsfonts}
\usepackage{bm}
\usepackage{graphicx}
\usepackage{color}

\begin{document}

\title{Kolmogorov-Hinze scales in turbulent superfluids}

\author{Tsuyoshi Kadokura}
\affiliation{Department of Engineering Science, University of
Electro-Communications, Tokyo 182-8585, Japan}

\author{Hiroki Saito}
\affiliation{Department of Engineering Science, University of
Electro-Communications, Tokyo 182-8585, Japan}

\date{\today}

\begin{abstract}
When a two-component mixture of immiscible fluids is stirred, the fluids
are split into smaller domains with more vigorous stirring.
We numerically investigate the sizes of such domains in a
fully-developed turbulent state of a two-component superfluid stirred
with energy input rate $\epsilon$.
For the strongly immiscible condition, the typical domain size is shown to
be proportional to $\epsilon^{-2/5}$, as predicted by the Kolmogorov-Hinze
theory in classical fluids.
For the weakly immiscible condition, quantum effects become pronounced and
the power changes from $-2 / 5$ to $-1 / 4$.
\end{abstract}

\maketitle

When oil is poured into water and these fluids are stirred, the oil becomes
split into droplets in the water.
The droplet sizes become smaller with more vigorous stirring.
Such disintegration phenomena in multicomponent fluids are ubiquitous in
nature, and are important in engineering and industry.

Kolmogorov~\cite{Kolmogorov49} and Hinze~\cite{Hinze} considered the
disintegration process of droplets, and estimated the size of droplets in
turbulent fluids.
In fully-developed turbulence, the energy is input into the system as
large-scale eddies, which cascades toward a smaller scale, resulting in
the Kolmogorov power law of the energy spectrum~\cite{Frisch}.
In such turbulent fluids, large-size droplets are unstable because they are
susceptible to deformation and disintegration due to the fluctuating
pressure of the surrounding fluid.
Small droplets are thus produced by the breakup of large droplets, and this
breakup process continues to a scale where the turbulent energy to break
up the droplets becomes balanced with the droplet energy that sustains
their shape.
Droplets smaller than this scale coalesce into large droplets.
Therefore, there exists a characteristic size $D$ for droplets in
turbulent fluids, which is referred to as the Kolmogorov-Hinze (KH)
scale, given by~\cite{Hinze}
\begin{equation} \label{Hinze}
D \sim (\sigma / \rho)^{3/5} \epsilon^{-2/5},
\end{equation}
where $\sigma$ is the interface tension coefficient, $\rho$ is the density
of the surrounding fluid, and $\epsilon$ is the energy input rate to
maintain the turbulence.
The KH scale has been experimentally verified in various systems~\cite{Clay,
  Shinnar, Sleicher, Arai, Deane}.
Furthermore, direct numerical simulations have been performed over the last
decade~\cite{Perlekar12, Skartlien, Perlekar14, Fan, Perlekar17, Rosti}.

In this Letter, we extend the study of KH scales to a quantum mechanical
system: the superfluid turbulence of a two-component Bose-Einstein
condensate (BEC).
We will show that the KH scale also appears in this superfluid system and
is modified by quantum effects.
Turbulent behavior in superfluids has been widely studied.
For single-component superfluids, a steady or decaying turbulent state
exhibits the Kolmogorov power law~\cite{Nore, Nore2, Stalp, Araki,
  Kobayashi, KobayashiJPSJ, Parker, Sasa, Baggaley, Leoni,
  Leoni2, Shukla, Muller, Polanco, Kobayashi21, Estrada}.
The turbulent behavior of gaseous BECs has also been experimentally
studied~\cite{Henn, Neely, Kwon}, and a power law behavior has been observed
recently~\cite{Thompson, Navon, Johnstone, Navon2, Galka, Garcia}.
A wide variety of systems have been studied theoretically, such as
two-dimensional systems~\cite{Nazarenko, Horng, Numasato, Bradley,
  Reeves, Muller20}, dipolar superfluids~\cite{Bland}, and boundary
layers~\cite{Stagg}.
Here, we focus on the turbulence in a two-component BEC.
Turbulence in multicomponent BECs has been investigated by many
researchers~\cite{Berloff, Takeuchi, Fujimoto, Fujimoto2, Fujimoto3,
  Fujimoto4, Fujimoto5, Tsubota, Vill, Kobyakov, Kang, Mithun, Das,
  Silva}.
In the context of domain-size scaling in multicomponent BECs, coarsening
dynamics following domain formation have been studied
extensively~\cite{Karl, Karl2, Kudo, Kudo2, Hofmann, De, Nicklas,
  Williamson, Williamson2, Williamson3, Bourges, Prufer, Fujimoto18,
  Takeuchi18, Symes}.
However, the KH scale, i.e., domain-size scaling in conjunction with
Kolmogorov turbulence, has not yet been investigated.

The KH scale in Eq.~(\ref{Hinze}) is derived as follows.
In a turbulent fluid, a domain undergoes fluctuating pressures that vary
over its size $D$, which causes deformation and disintegration of the
domain.
This pressure difference can be expressed as $\sim \rho v^2 \equiv
P_{\rm turbulence}$, where $v$ is the velocity difference of the
surrounding fluid over a size $D$.
Within the inertial range of an isotropic homogeneous turbulence,
the statistical average of $v^2$ obeys the Kolmogorov two-thirds
law~\cite{Frisch}, $\bar{v^2} \propto (\epsilon D)^{2/3}$, and hence
$P_{\rm turbulence} \sim \rho (\epsilon D)^{2/3}$.
On the other hand, a domain tends to sustain its shape and resist
disintegration.
This sustaining force arises from the interface tension, and the pressure
required to deform the domain is estimated to be $\sim \sigma / D
\equiv P_{\rm sustain}$~\cite{Landau}. 
Breakup of domains to smaller sizes stops at the scale that satisfies
\begin{equation} \label{pp}
  P_{\rm sustain} \sim P_{\rm turbulence},
\end{equation}  
which gives the KH scale in Eq.~(\ref{Hinze}).

For an immiscible two-component BEC, the interface tension, which arises
from the interatomic interaction and quantum pressure, is well-defined, as
in classical fluids~\cite{Ao, Barankov, Schae}.
Therefore, we expect that the KH scale in Eq.~(\ref{Hinze}) also emerges in
two-component BECs, when the thickness of the interface $W$ is much
smaller than the domain size $D$.
However, when $W$ is comparable to or larger than $D$, the picture of the
interface tension breaks down in the derivation of Eq.~(\ref{Hinze}).
The interface thickness $W$ is determined by the competition between the
quantum pressure and the intercomponent repulsion, and $W$ becomes large
when the former dominates the latter.
Thus, in the limit of large $W$, the mechanism to sustain domains against
disintegration originates mainly from the quantum pressure $P_{\rm sustain}
\sim n \hbar^2 / (m D^2)$ instead of $P_{\rm sustain} \sim \sigma /
D$, where $n$ and $m$ are the atomic number density and mass,
respectively.
In this case, Eq.~(\ref{pp}) gives the characteristic size as
\begin{equation} \label{qHinze}
D \sim (\hbar / m)^{3/4} \epsilon^{-1/4}.
\end{equation}
Therefore, in the limit of weak segregation with large $W$, the quantum
mechanical effect becomes pronounced and the KH scale is expected to change
from the $-2/5$ to $-1/4$ power law with respect to $\epsilon$.
In the remainder of this Letter, we will corroborate this prediction using
numerical simulations of the coupled Gross-Pitaevskii (GP) equations.

In the mean-field approximation, a two-component BEC at zero
temperature is described by the coupled GP equations,
\begin{subequations}
\label{eq:BIN}
\begin{eqnarray}
i \hbar \frac{\partial \psi_1}{\partial t} & = & 
\left(
-\frac{\hbar^2}{2m}\nabla^2 + V_{\textrm{ext}}
+ g_{11}|\psi_1|^2 + g_{12}|\psi_2|^2
\right) \psi_1,
\nonumber \\
\\
i \hbar \frac{\partial \psi_2}{\partial t} & = & 
\left(
-\frac{\hbar^2}{2m}\nabla^2 + V_{\textrm{ext}}
+ g_{22}|\psi_2|^2 + g_{12}|\psi_1|^2
\right) \psi_2,
\nonumber \\
\end{eqnarray}
\end{subequations}
where $\psi_j(\bm{r}, t)$ is the macroscopic wave function for the
$j$th component, $V_{\rm ext}(\bm{r}, t)$ is the external stirring
potential, and $g_{jj'} = 4 \pi \hbar^2 a_{jj'} / m$ with $a_{jj'}$
being the $s$-wave scattering length between the $j$th and $j'$th
components.

The miscibility between the two components is determined by the coupling
coefficients $g_{jj'}$.
The two superfluids are immiscible and phase separation occurs when
$g_{12}^2 > g_{11} g_{22}$ is satisfied~\cite{Pethick}.
In the following, for simplicity, we assume $g_{11} = g_{22} \equiv g
> 0$ and $g_{12} > 0$;
therefore, the immiscible condition reduces to $g_{12} > g$.
The phase separation of immiscible components produces an interface, at
which excess energy arises, resulting in interface tension.
For $g_{12} / g - 1 \ll 1$, the interface tension coefficient is
given by~\cite{Ao, Barankov, Schae}
\begin{equation} \label{sigma}
\sigma \simeq \left[ \frac{\hbar^2 n^3}{2m} (g_{12} - g)
  \right]^{1/2}.
\end{equation}
The interface thickness $W$ between two components, over which the
density of each component changes from 0 to $n$ (or $n$ to 0), has the
form
\begin{equation} \label{W}
W \simeq \xi (g_{12} / g - 1)^{-1/2},
\end{equation}
where $\xi \equiv \hbar / (mgn)^{1/2}$ is the healing length.

In the following, the length, time, and wave functions are normalized
by $W$, $m W^2 / \hbar$, and $\sqrt{n}$, respectively, where $W$ is
the interface thickness defined in Eq.~(\ref{W}) and $n$ is the
average density of each component.
In this unit, the normalized interaction coefficients, $\tilde g$ and
$\tilde g_{12}$, in the GP equation become~\cite{SM}
\begin{equation} \label{g}
\tilde g = \frac{g}{g_{12} - g}
\end{equation}
and $\tilde g_{12} = g_{12} / (g_{12} - g) = \tilde g + 1$; therefore,
the interaction coefficients are reduced to the single parameter
$\tilde g$.
The GP equation is numerically solved using the split-step Fourier
method~\cite{Recipe}.
We consider a box of size $L^3 = (256 \tilde \xi)^3$ with a periodic
boundary condition, where the nondimensional healing length is
$\tilde\xi \equiv \xi / W = 1 / \sqrt{\tilde g}$.
The box is discretized into a $512^3$ mesh, and the spatial resolution
is 0.5 $\tilde \xi$.
The two components are equally populated, $\int |\psi_1|^2 d\bm{r} =
\int |\psi_2|^2 d\bm{r} = L^3$, and the initial state has a uniform
density with random phases on each mesh.

To input the large-scale turbulent energy, the system is stirred using
plate-shaped potentials given by
\begin{equation} \label{Vext}
  V_{\textrm{ext}}(\bm{r}, t) = V_0 \sum_{\{X, Y\}} e^{-[X - \frac{L}{2}
      \sin(\Omega t + \phi_{XY})]^2} \theta(L/4 - |Y|), 
\end{equation}
where the potential height is taken to be $V_0 = 2$, $\theta$ is the
Heaviside step function, and the summation is taken over $\{X, Y\} =
\{x, y\}$, $\{y, z\}$, and $\{z, x\}$ with $\phi_{XY} = 0$, $2\pi /
3$, and $4\pi / 3$, respectively.
The three plate-shaped potentials oscillating in the $x$, $y$, and $z$
directions produce isotropic turbulence.
The maximum Mach number of the plate-shaped potentials is defined as
\begin{equation} \label{Mach}
M \equiv \frac{\Omega L}{2 v_s},
\end{equation}
where $v_s = \sqrt{2\tilde g}$ is the sound velocity.

To realize a steady turbulent state, the energy must be dissipated on
a small length scale.
For this purpose, the term $-\gamma (\nabla \cdot \bm{J}) \psi_j$ is
added to the right-hand sides of Eq.~(\ref{eq:BIN}), where $\gamma$ is
a positive constant and
\begin{equation} \label{J}
  \bm{J} = \frac{1}{2i} \sum_{j=1}^2 \left( \psi_j^* \nabla \psi_j -
  \psi_j \nabla \psi_j^* \right).
\end{equation}
This phenomenological dissipation term mimics the viscous term in the
Navier-Stokes equation and reduces the energy of the system while
maintaining the unitarity~\cite{SM}.
The value of $\gamma$ is selected in such a way that energy
dissipation occurs predominantly on a scale below those for the inertial
range and the domain size.
The larger-scale dynamics are not affected by the details of the
dissipation, as long as it occurs on a sufficiently small
scale~\cite{SM}.

\begin{figure}[t]
\includegraphics[width=8.0cm]{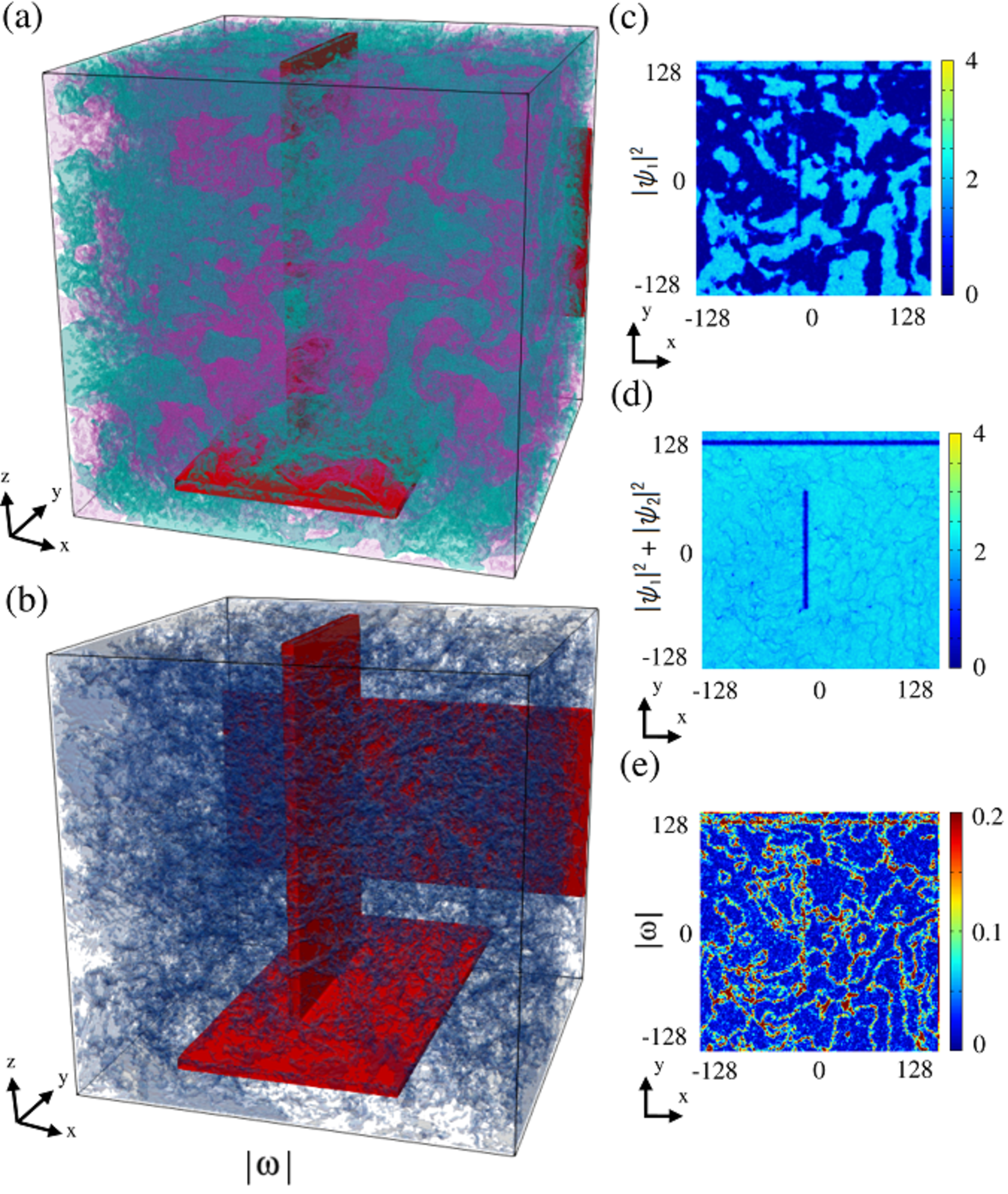}
\caption{
Snapshots of a fully-developed turbulent state at $t = 2000$ for
$\tilde g \equiv g / (g_{12} - g) = 6.25$ and $M = 0.18$.
A two-component BEC is stirred by plate-shaped potentials (red or
black in (a) and (b)) to generate a turbulent state.
The three plates oscillate in the $x$, $y$, and $z$ directions
(perpendicular to the surface) from end to end in the box.
(a) Isodensity surfaces of $|\psi_1|^2$ (purple or dark gray) and
$|\psi_2|^2$ (green or light gray).
(b) Isodensity surfaces of $|\bm{\omega}|$.
(c), (d), and (e) Cross-sectional views ($z = 0$) of $|\psi_1|^2$,
$|\psi_1|^2 + |\psi_2|^2$, and $|\bm{\omega}|$, respectively.
}
\label{f:setup}
\end{figure}
Figure~\ref{f:setup}(a) shows isodensity surfaces of $|\psi_1|^2$ and
$|\psi_2|^2$ after the fully-developed turbulent state is achieved.
The two components are separated and domains are formed in each
component because of the immiscible condition $g_{12} > g$.
The domain sizes in Fig.~\ref{f:setup}(a) are typically $\sim 10$
(note that the interface thickness $W$ is unity in the present unit),
and thus the KH scale is expected to be in the region of
Eq.~(\ref{Hinze}) (rather than Eq.~(\ref{qHinze})),
which will be investigated later.

Figures~\ref{f:setup}(c) and \ref{f:setup}(d) show cross-sectional views of
the densities $|\psi_1|^2$ and $|\psi_1|^2 + |\psi_2|^2$, respectively.
Although $|\psi_1|^2$ (or $|\psi_2|^2$) largely varies in space due to
the phase separation (Fig.~\ref{f:setup}(c)), the total density far
from the stirring potentials is almost uniform (Fig.~\ref{f:setup}(d))
and density holes arising from quantized vortices are rarely observed,
since the velocity of the stirring potential is much lower than the
sound velocity of the density waves.
This situation is different from the quantum turbulence in a
single-component system, in which quantized vortices play a central
role in the energy cascade.
This difference arises because the vorticity of the mass current is
not quantized in the two-component system.
To observe this, we define the vorticity of the mass-current velocity as
$\bm{\omega} = \nabla \times [\bm{J} / (|\psi_1|^2 + |\psi_2|^2)]$.
Figures~\ref{f:setup}(b) and \ref{f:setup}(e) show the distribution of
$|\bm{\omega}|$.
It is evident from Figs.~\ref{f:setup}(c) and \ref{f:setup}(e) that
the vorticity is localized around the interfaces of the domains.

\begin{figure}[t]
\includegraphics[width=7.5cm]{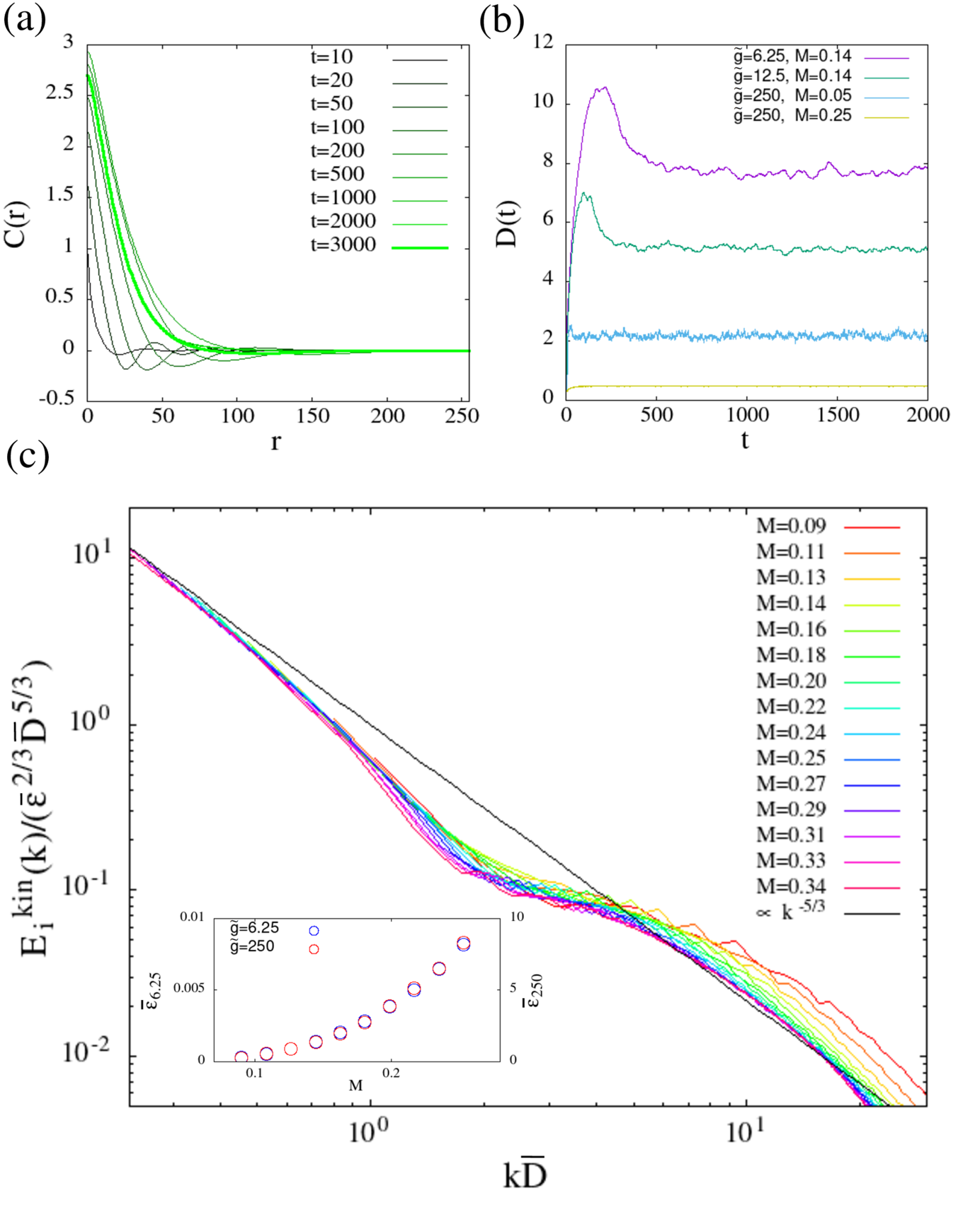}
\caption{
(a) Time development of the density correlation function $C(r)$
  defined in Eq.~(\ref{corr}) for $\tilde{g} = 12.5$ and $M = 0.14$.
(b) Time development of the typical domain size $D$ (full width at
  half maximum of $C(r)$) for various values of $\tilde g$ and $M$.
(c) Incompressible kinetic energy spectra $E^{\rm kin}_i(k)$ for
  $\tilde{g} = 6.25$ and various values of $M$~\cite{SM}, where
  the time average is taken over two stirring periods.
  $E^{\rm kin}_i(k)$ is compensated by $\bar\epsilon^{2/3}$ and the
  length is rescaled by $\bar D$, where $\bar\epsilon$ and $\bar D$
  are time-averaged steady values of the energy input rate and the
  typical domain size, respectively.
  The straight line represents $k^{-5/3}$.
  The inset shows $\bar\epsilon$ as a function of $M$ for $\tilde g =
  6.25$ and 250.
}
\label{f:arrest}
\end{figure}
The typical size of the domains can be evaluated from the density
correlation function,
\begin{equation} \label{corr}
  C(r) = \langle d(\bm{r}') d(\bm{r}' + \bm{r}) \rangle,
\end{equation}
where $d = |\psi_1|^2 - |\psi_2|^2$ is the density-imbalance
distribution and $\langle \cdots \rangle$ represents the average over
the position $\bm{r}'$ and the direction of $\bm{r}$.
Figure~\ref{f:arrest}(a) shows the time evolution of $C(r)$.
Since the initial wave function has a random distribution, $C(r)$ is
initially narrow, which then becomes broader and reaches a steady
shape for $t \gtrsim 1000$.
We have confirmed that the steady shape of $C(r)$ is not dependent on
the initial state.
We define the typical domain size $D$ as the full width at half
maximum of the correlation function $C(r)$.
Figure~\ref{f:arrest}(b) shows the time development of $D$ for
different values of $\tilde g$ and $M$.
The size $D$ decreases with increasing $\tilde g$ and $M$.

The energy input rate per atom is obtained by
\begin{equation}
  \epsilon = \frac{\int (|\psi_1|^2 + |\psi_2|^2) \dot{V}_{\rm ext}
    d\bm{r}}{\int (|\psi_1|^2 + |\psi_2|^2) d\bm{r}},
\end{equation}
where the time dependence of the potential $V_{\rm ext}$ is given in
Eq.~(\ref{Vext}).
The value of $\epsilon$ (and also $D$) fluctuates over time due to the
random nature of the turbulence.
The sinusoidal motion of the plate-shaped potentials also causes periodic
fluctuation.
Therefore, we take the temporal average of these quantities, $\bar\epsilon$
and $\bar D$, over a sufficiently long time after the steady turbulent state
is achieved.
The inset in Fig.~\ref{f:arrest}(c) shows $\bar\epsilon$ as a function of
$M$.

To confirm that the system has reached the Kolmogorov turbulence
state, we calculate the kinetic energy spectrum $E^{\rm kin}_i(k)$ of
the incompressible velocity field of the mass current~\cite{SM}, which
is shown in Fig.~\ref{f:setup}(c).
Since the Kolmogorov theory predicts $E^{\rm kin}_i(k) \propto
\bar{\epsilon}^{2/3} k^{-5/3}$, the plots in Fig.~\ref{f:arrest}(c)
are compensated by $\bar{\epsilon}^{2/3}$.
The length is also rescaled by the domain size $\bar{D}$ to observe
the effect of the domains on the energy spectrum.
Figure~\ref{f:arrest}(c) shows that the lines of the energy spectra
with different $\bar{\epsilon}$ and $\bar{D}$ collapse into a single
universal line with a slope of $\simeq -5 / 3$ on a scale larger than
the domain size ($k \bar{D} \lesssim 1$), which implies that the
Kolmogorov energy cascade occurs on this scale.
At the scale of $k \bar{D} \sim 1$, the energy cascade is arrested by
the domains~\cite{Perlekar14}, which results in a ``bump'' in the
energy spectrum, as shown in Fig.~\ref{f:arrest}(c).
This situation is similar to the case of a single-component system, in
which the inertial range is terminated at the scale of the mean
distance between quantized vortices~\cite{Nore, Lvov}.

\begin{figure}[t]
\includegraphics[width=7.5cm]{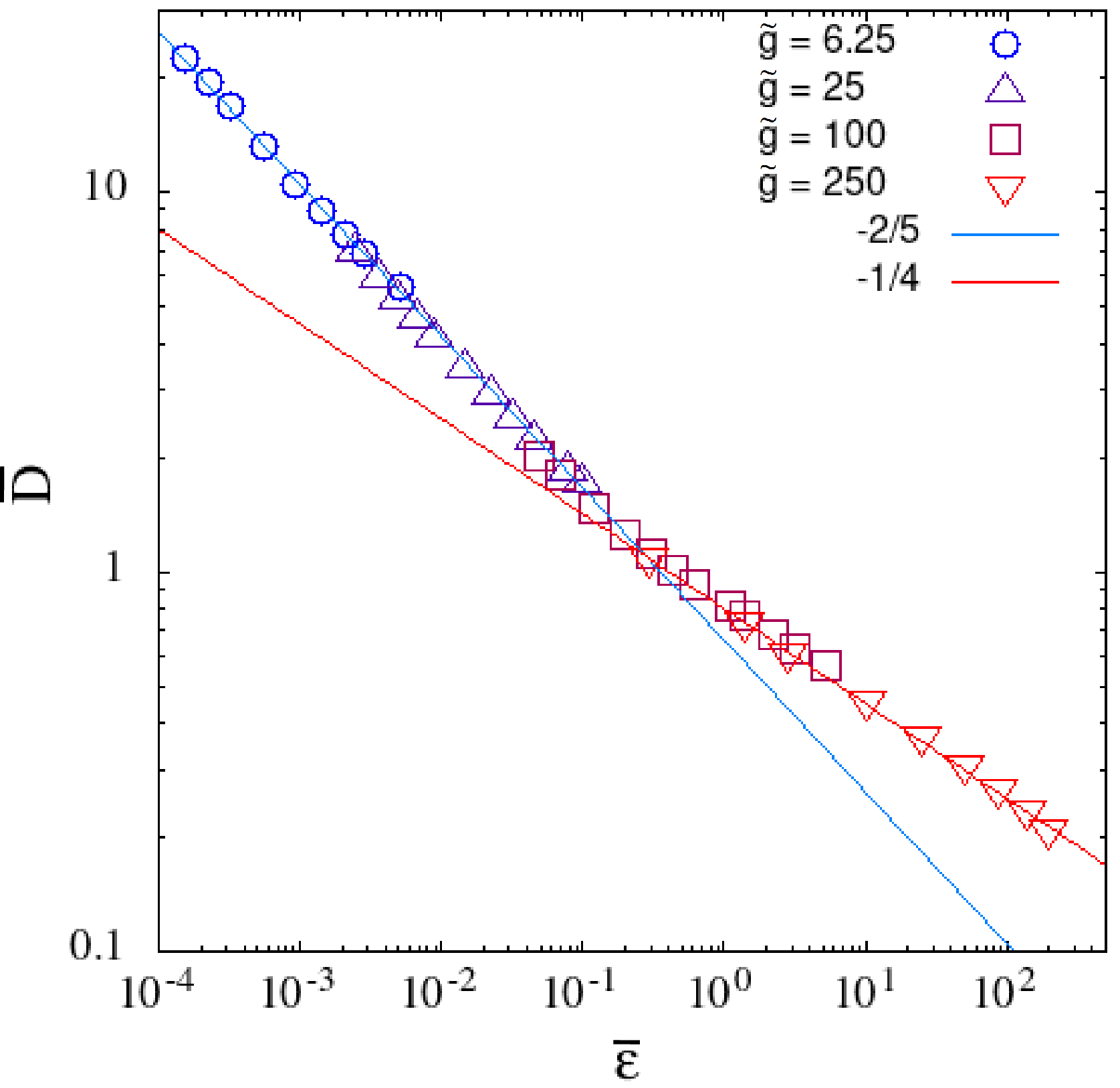}
\caption{
Typical domain size $\bar D$ versus energy input rate $\bar\epsilon$
for the steady turbulent state with $\tilde{g} = 6.25$, 25,
100, and 250.
For each value of $\tilde g$, $M$ is varied within an appropriate
range (see text).
The slopes of the lines are $-2/5$ and $-1/4$ for comparison with
Eqs.~(\ref{Hinze}) and (\ref{qHinze}).
Note that $\bar D$ is normalized by the interface width $W$ in the
present unit, and the crossover between the two power laws occurs at
$\bar D \sim W$.
}
\label{f:hinze}
\end{figure}
Now we are ready to investigate the KH scales in a turbulent superfluid in
the classical and quantum regimes, as given in Eqs.~(\ref{Hinze}) and
(\ref{qHinze}), respectively.
The results are shown in Fig.~\ref{f:hinze}, which are the main
results obtained in this study.
Figure~\ref{f:hinze} plots the typical domain size $\bar D$ versus the
energy input rate $\bar\epsilon$ for various values of the Mach number
of the stirring potential $M$ and the normalized interaction
coefficient $\tilde g$.
For $\bar D \gg 1$ (note that $\bar D$ is normalized by the interface
width $W$), the plots obey the power law $\propto
\bar\epsilon^{-2/5}$, which agrees with the classical KH scale in
Eq.~(\ref{Hinze}).
This implies that the two components are well separated and the
mechanism that sustains the domains against disintegration can be
described by the interface tension in this region.
For $\bar D \ll 1$, on the other hand, the plots in Fig.~\ref{f:hinze}
follow the power law $\propto \bar\epsilon^{-1/4}$, which agrees with
the KH scale in the quantum region in Eq.~(\ref{qHinze}) and implies
that the mechanism that sustains domains is mainly the quantum kinetic
pressure arising from the uncertainty principle.

In the numerical simulations in Fig.~\ref{f:hinze}, the plot range for
each $\tilde g$ is restricted, because the domain size $\bar D$ is
limited by the size of the numerical box, and the energy input rate
$\bar\epsilon$ is limited by the maximum velocity allowed for the
plate-shaped potentials.
In the present normalization, the box size is $L = 256 \tilde \xi =
256 / \sqrt{\tilde g}$, and hence $\tilde D$ can be larger for smaller
$\tilde g$ (left-hand plots in Fig.~\ref{f:hinze}).
On the other hand, the Mach number $M$ of the plate-shaped potentials
must be smaller than about unity, or the total density would be
significantly disturbed and the present picture (domains formed by
phase separation) breaks down.
In the present unit, the sound velocity is $v_s = \sqrt{2\tilde g}$;
therefore, we can drive the stirring potential faster for larger
$\tilde g$.
This is the reason why the energy input rate $\bar\epsilon$ can be
made larger for larger $\tilde g$, and the more rightward region can
be plotted in Fig.~\ref{f:hinze}.
Thus, although $\bar D$ and $\bar\epsilon$ are restricted to narrow
ranges for each value of $\tilde g$ in the present numerical
simulations, the plots in Fig.~\ref{f:hinze} can be extended to a wide
range, which corroborates the existence of the two power laws in the
superfluid KH scale.

Finally, we discuss the possible experimental realization of the present
results.
A box potential would be suitable to avoid complexity arising from the
inhomogeneous $|\psi_1|^2 + |\psi_2|^2$ distribution in a harmonic
potential.
The stirring potential can be produced by a far-off-resonance laser
beam.
Shaking of an optical box can also be used to generate the turbulent
state~\cite{Navon}.
The typical size of the domains can be inferred from the imaging data,
where slice imaging of a three-dimensional distribution may be
required~\cite{Andrews}.
It is difficult to measure the energy input rate directly; therefore, the
support of numerical simulation is necessary, which provides the relation
between the motion of the potential and the energy input rate, as in the
inset in Fig.~\ref{f:arrest}(c).
The interaction $g_{12}$ can be varied using the Feshbach resonance
technique.

In conclusion, we have investigated the KH scale of domain sizes in
immiscible two-component superfluids in a fully-developed turbulent state.
We predict that two regions of the KH scale exist with different power
laws, which reflect the quantum properties of the system.
Numerical simulations of the coupled GP equations were performed, and the
typical domain size $\bar D$ was confirmed to obey the power laws with
respect to the energy input rate $\bar\epsilon$.
The power changes from $-2/5$ to $-1/4$ with increasing
$\bar\epsilon$, and the crossover between these classical and quantum
KH scales is located at the region where $\bar D$ is comparable to the
interface thickness $W$.
A possible extension of this study is a three-component system, in
which the third component can change the interface tension of the
other two components~\cite{Jimbo}, resulting in emulsification.

This work was supported by JSPS KAKENHI Grant Numbers JP20K03804 and
JP23K03276.

\newpage

\begin{center}
{\bf Supplemental Material}
\end{center}

\section{Energy dissipation}

We must include phenomenological energy dissipation in the GP equation
to study the steady turbulence, in which energy is continuously input
into the system.
The energy dissipation should predominantly occur on a small length
scale, and should not affect the large-scale dynamics.
In previous studies of steady quantum
turbulence~\cite{Kobayashi, KobayashiJPSJ}, nonunitary energy
dissipation was used, where the GP equation was represented in
wave-number space and $i$ on the left-hand side was replaced with $i -
\Gamma$ only for large wave numbers, where $\Gamma$ is a positive
constant.
To recover the unitarity, the term $-\mu(t) \psi$ is added to the
right-hand side of the GP equation and $\mu(t)$ is chosen in such a
way that the norm of the wave function is kept constant, which
corresponds to renormalization of the wave function at each time.
However, this prescription is nonlocal in wave-number space, since
the reduction of the norm for a large wave number is compensated by the
whole wave numbers.
This prescription is also nonlocal in real space for a similar
reason.

\begin{figure}[t]
\includegraphics[width=7.5cm]{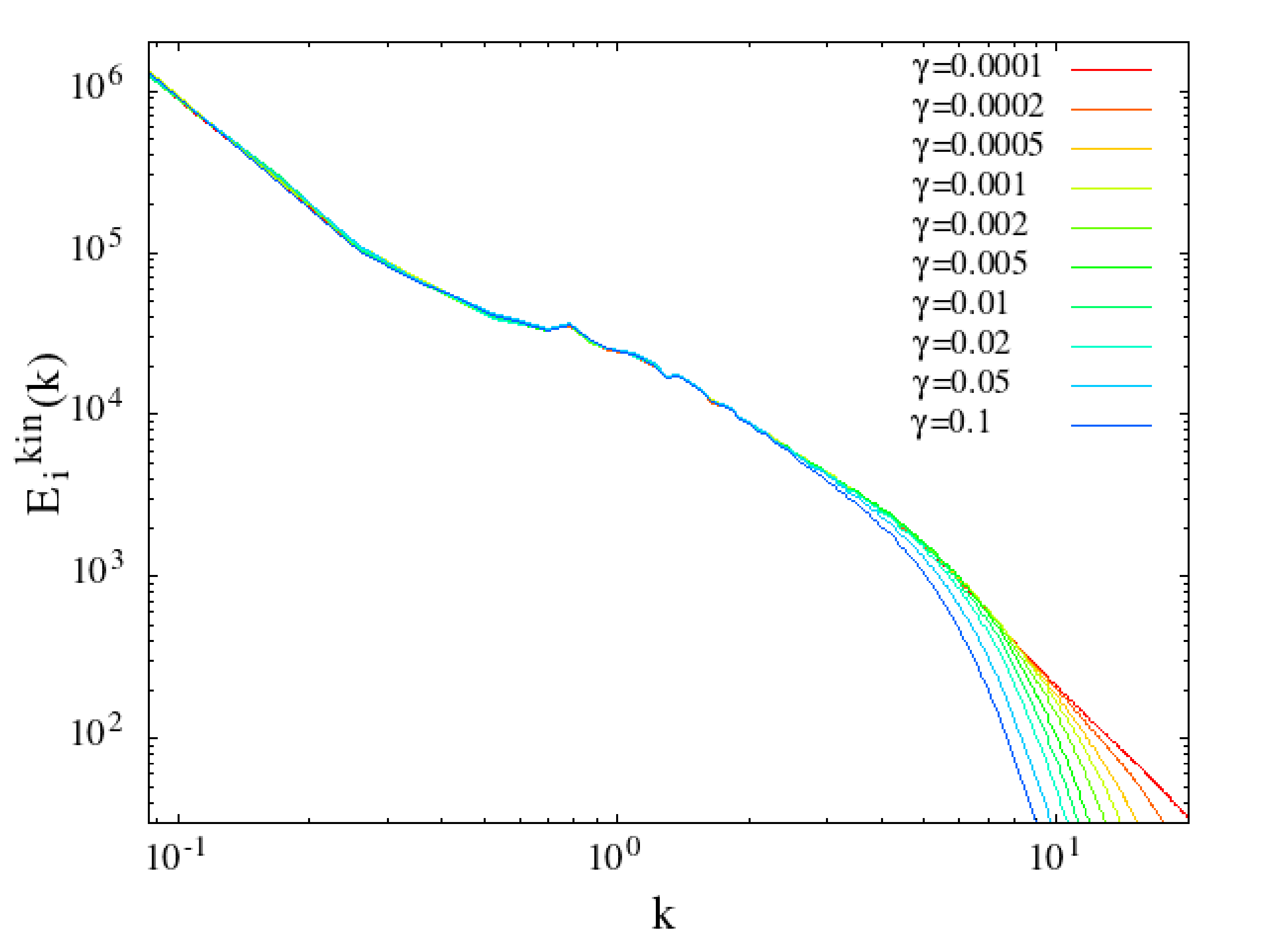}
\caption{
  $\gamma$ dependence of incompressible kinetic energy spectrum
  $E_i^{\rm kin}(k)$ of steady turbulent state for $\tilde g = 12.5$
  and $M = 0.14$.
}
\label{f:gamma}
\end{figure}
\begin{figure*}[tb]
\includegraphics[width=15cm]{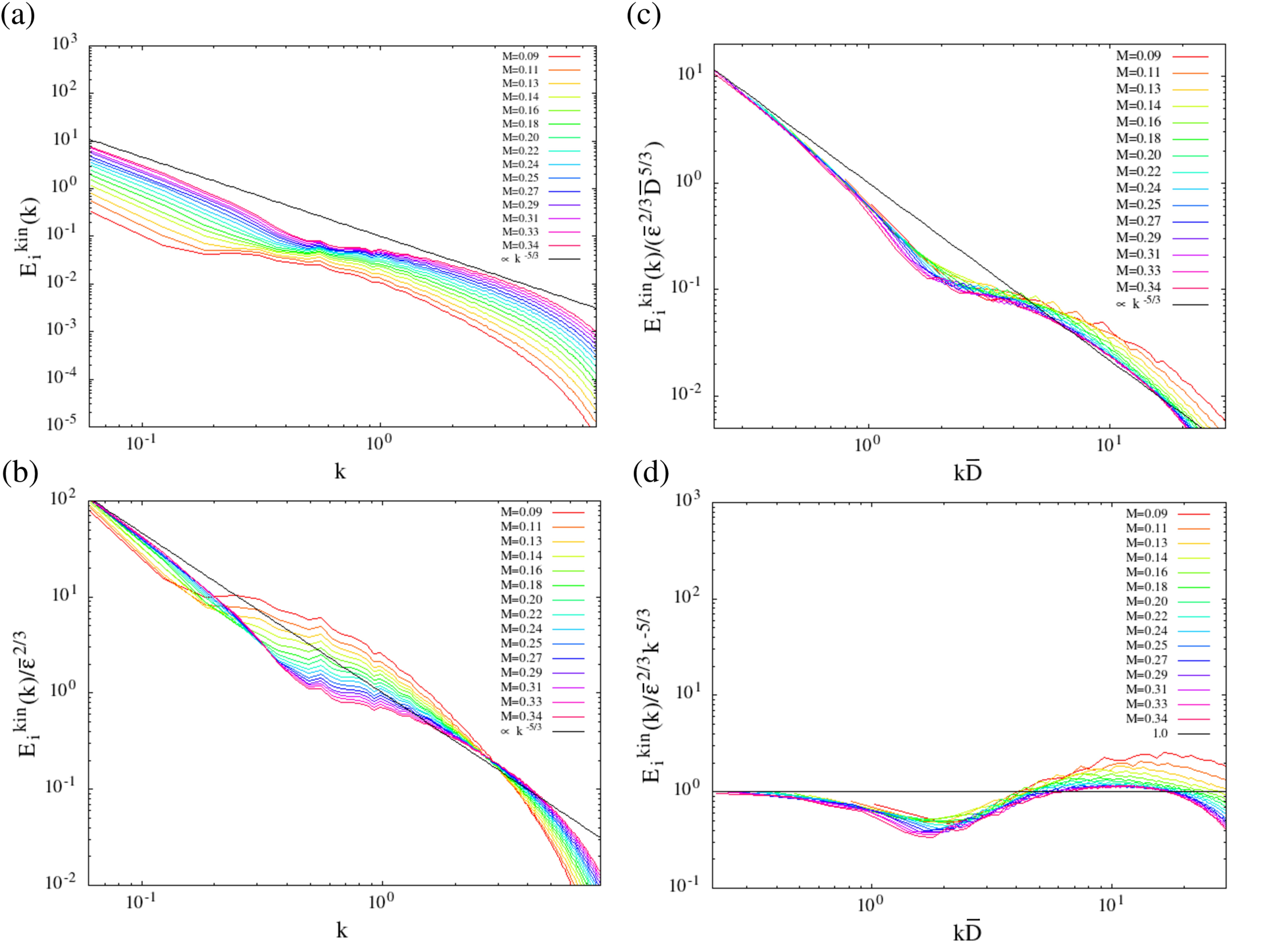}
\caption{
Incompressible kinetic energy spectra $E_i^{\rm kin}(k)$ for $\tilde g
= 6.25$ with various values of the Mach number $M$ of the stirring
potentials.
(a) $E_i^{\rm kin}(k)$ versus $k$,
(b) $E_i^{\rm kin}(k) / \bar{\epsilon}^{2/3}$ versus $k$,
(c) $E_i^{\rm kin}(k) / (\bar{\epsilon}^{2/3} \bar D^{5/3})$ versus
$k\bar D$,
and (d) $E_i^{\rm kin}(k) / (\bar{\epsilon}^{2/3} k^{-5/3})$ versus
$k\bar D$,
where $\bar{\epsilon}$ and $\bar D$ are the time-averaged
energy-input rate and the domain size, respectively.
Panel (c) is the same as Fig.~2(c) in the main text.
}
\label{f:ek}
\end{figure*}
Here, we propose another phenomenological method for energy
dissipation that assures unitarity and locality.
The term $-\gamma (\nabla \cdot \bm{J}) \psi_j$ is added to the
right-hand side of the GP equation, where $\gamma$ is a positive
constant and
\begin{equation}
  \bm{J} = \frac{\hbar}{2mi} \sum_{j=1}^2 \left( \psi_j^* \nabla
  \psi_j - \psi_j \nabla \psi_j^* \right).
\end{equation}
Intuitively, when $\nabla \cdot \bm{J}$ is negative (positive), i.e.,
the time derivative of the density is positive (negative), the added
term serves as a positive (negative) potential, which counteracts the
inflow (outflow) and results in a reduction of the compressible
kinetic energy.
Since the added term is proportional to the second power of the wave
numbers, it only affects large wave numbers for a sufficiently
small $\gamma$.
It should be noted that the large-scale phenomena are not dependent on
the details of the dissipation, as long as the dissipation occurs only
on a small length scale.

Figure~\ref{f:gamma} shows incompressible kinetic energy spectra
$E_i^{\rm kin}(k)$ (definition given in Sec.~\ref{s3}) of the steady
turbulent state for various values of $\gamma$.
The energy spectra $E_i^{\rm kin}(k)$ rapidly decays on the small
scale ($k \gtrsim 10$) depending on the value of $\gamma$, while
$E_i^{\rm kin}(k)$ is not dependent on the value of $\gamma$ at larger
scales ($k \lesssim 1$).
This indicates that the energy dissipation only occurs on the small
scale and does not affect the large scale.
In the main text, $\gamma = 0.005$ is adopted.

The dissipation term $-\gamma (\nabla \cdot \bm{J}) \psi_j$ mimics the
viscous term in the classical Navier-Stokes equation.
To describe this, for simplicity, let us consider the single-component
GP equation,
\begin{equation} \label{GP}
  i\hbar \frac{\partial\psi}{\partial t}
  = -\frac{\hbar^2}{2m} \nabla^2 \psi + V \psi + g |\psi|^2 \psi -
  \gamma (\nabla \cdot \bm{J}) \psi.
\end{equation}
Using the Madelung transformation $\psi = \sqrt{n} e^{i\theta}$, the
real part of Eq.~(\ref{GP}) becomes
\begin{equation} \label{NS}
  \frac{\partial\bm{v}}{\partial t} + (\bm{v} \cdot \nabla) \bm{v}
  = -\frac{1}{m} \nabla U + \frac{\gamma}{m}
  \left[ n \nabla^2 \bm{v} + \nabla (\bm{v} \cdot \nabla n) \right],
\end{equation}
where $\bm{v} = \frac{\hbar}{m} \nabla\theta$ and $U =
-\frac{\hbar^2}{2m\sqrt{n}} \nabla^2 \sqrt{n} + V + gn$.
Assuming that the density $n$ is almost uniform, the first term is
dominant in the square bracket, which has the same form as the viscous
term in the Navier-Stokes equation with a kinematic viscosity $\nu =
\gamma n / m$.

\section{Normalization of the Gross-Pitaevskii equation}

The Gross-Pitaevskii (GP) equation that we solve has the form,
\begin{eqnarray}
  i \hbar \frac{\partial \psi_j}{\partial t} & = & -\frac{\hbar^2}{2m}
  \nabla^2 \psi_j + V_{\rm ext} \psi_j + g |\psi_j|^2 \psi_j
  + g_{jj'} |\psi_{j'}|^2 \psi_j
  \nonumber \\
  & &  - \gamma (\nabla \cdot \bm{J}) \psi_j,
\end{eqnarray}
where $(j, j') = (1, 2)$ and $(2, 1)$.
The nondimensional quantities are introduced as $\tilde{\bm{r}} =
\bm{r} / {\cal L}$, $\tilde t = \hbar t / (m {\cal L}^2)$, and
$\tilde{\psi}_j = \psi_j / \sqrt{n}$, where ${\cal L}$ and
$n$ are the unit length and density, respectively.
Substituting these into the GP equation, we obtain
\begin{eqnarray}
  i \frac{\partial \tilde{\psi}_j}{\partial \tilde t} & = & -\frac{1}{2}
  \tilde\nabla^2 \tilde{\psi}_j + \tilde{V}_{\rm ext} \tilde{\psi}_j
  + \tilde{g} |\tilde{\psi}_j|^2 \tilde{\psi}_j
  + \tilde{g}_{12} |\tilde{\psi}_{j'}|^2 \tilde{\psi}_j
  \nonumber \\
  & &  - \tilde{\gamma} (\tilde\nabla \cdot \tilde{\bm{J}})
  \tilde{\psi}_j,
\end{eqnarray}
where $\tilde{V}_{\rm ext} = m {\cal L}^2 V_{\rm ext} / \hbar^2$,
$\tilde g = m {\cal L}^2 g n / \hbar^2$,
$\tilde g_{12} = m {\cal L}^2 g_{12} n / \hbar^2$,
$\tilde\gamma = n \gamma / \hbar$,
and $\tilde{\bm{J}} = \sum_j (\tilde{\psi}_j^* \tilde\nabla
\tilde{\psi}_j - \tilde{\psi}_j \tilde\nabla \tilde{\psi}_j^*) / (2i)$.

Here, the length unit ${\cal L}$ is taken to be the interface width $W
= \hbar [m n (g_{12} - g)]^{-1/2}$, and we have $\tilde g = g /
(g_{12} - g)$ and $\tilde{g}_{12} = g_{12} / (g_{12} - g) = \tilde g +
1$.
In the main text, the tildes in the nondimensional quantities are
omitted, except $\tilde g$.

\section{Kinetic energy spectra}
\label{s3}

\begin{figure}[tb]
\includegraphics[width=7.5cm]{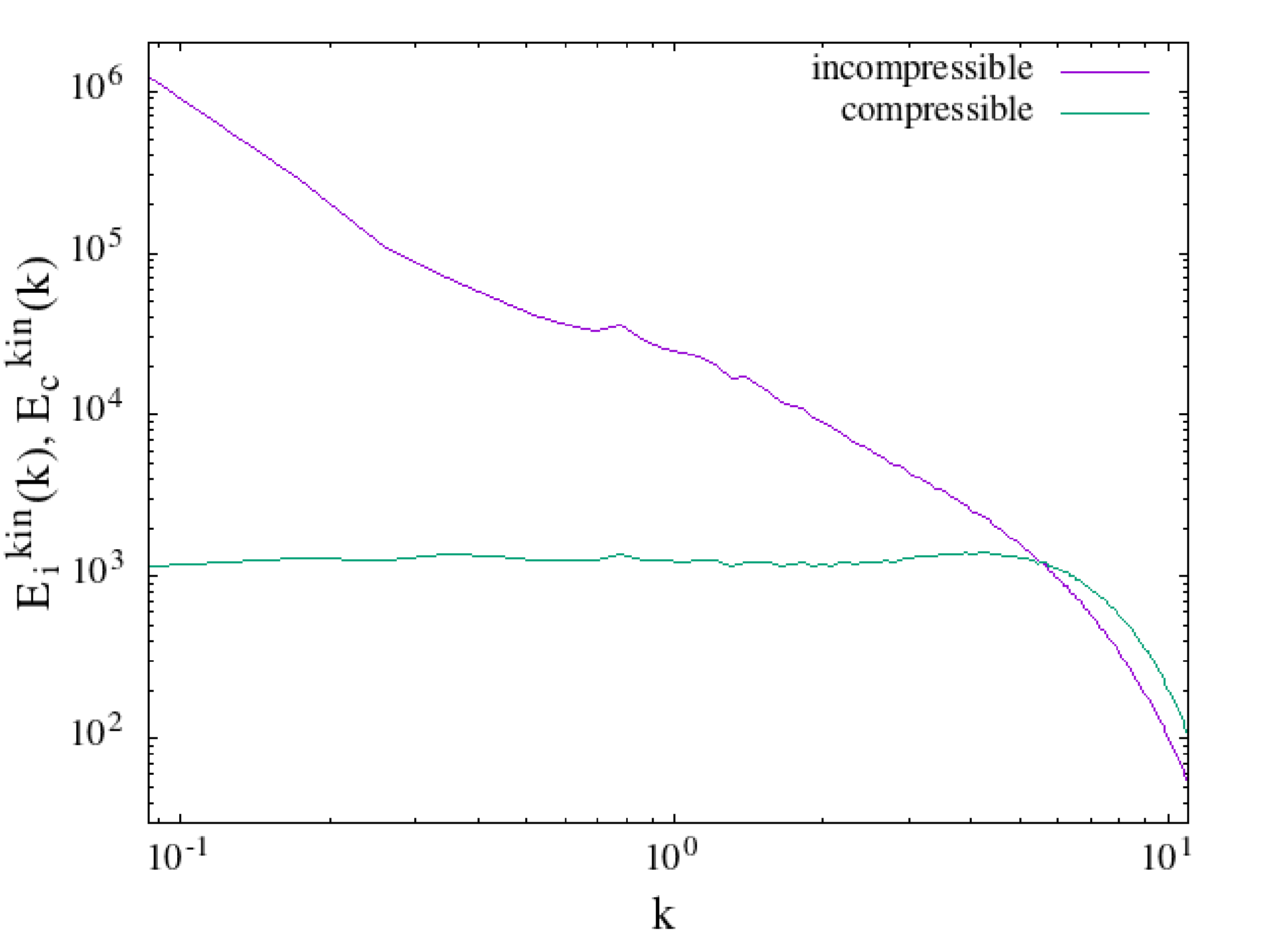}
\caption{
  Compressible and incompressible kinetic energy spectra,
  $E_c^{\rm kin}(k)$ and $E_i^{\rm kin}(k)$, of the steady turbulent
  state for $\tilde g = 12.5$ and $M = 0.14$.
  The time average is taken over two stirring periods.
}
\label{f:ec}
\end{figure}
In a manner similar to the single-component case, we consider the
kinetic energy of the mass current as
\begin{equation} \label{Ekin}
  {\cal E}^{\rm kin} = \frac{1}{2} \int d{\bm{r}} \rho \bm{v}^2 
  = \frac{1}{2} \int \frac{d\bm{k}}{(2\pi)^3} |\bm{W}(\bm{k})|^2,
\end{equation}
where $\rho = (|\psi_1|^2 + |\psi_2|^2) / \int (|\psi_1|^2 +
|\psi_2|^2) d\bm{r}$ is the density per atom and $\bm{v} = \bm{J} /
(|\psi_1|^2 + |\psi_2|^2)$ is the velocity field of the mass current.
In Eq.~(\ref{Ekin}), the Fourier component is defined as
$\bm{W}(\bm{k}) = \int d\bm{r} \sqrt{\rho} \bm{v}
e^{-i \bm{k} \cdot \bm{r}}$.
The kinetic energy ${\cal E}^{\rm kin}$ can be decomposed into
compressible and incompressible parts as
\begin{equation}
  {\cal E}^{\rm kin} = \frac{1}{2} \int \frac{d\bm{k}}{(2\pi)^3} \left[
  |\bm{W}_c(\bm{k})|^2 + |\bm{W}_i(\bm{k})|^2 \right],
\end{equation}
where $\bm{W}_c = (\bm{W} \cdot \bm{k}) \bm{k} / k^2$
and $\bm{W}_i = \bm{W} - \bm{W}_c$.
The compressible and incompressible kinetic energy spectra are defined
by
\begin{equation}
  E^{\rm kin}_{c, i}(k) = \frac{1}{2} \int
  \frac{k^2 \sin\theta_k d\theta_k d\phi_k}{(2\pi)^3}
  |\bm{W}_{c, i}(\bm{k})|^2,
\end{equation}
where $\theta_k$ and $\phi_k$ are the polar and azimuthal angles in
wave-number space, respectively.

Figure~\ref{f:ek} shows the energy spectra for the steady turbulent
state for $\tilde g = 6.25$ and various values of the Mach number $M$,
where the time average is taken for two stirring periods after steady
turbulence is achieved.
The raw data for $E_i^{\rm kin}(k)$ are shown in Fig.~\ref{f:ek}(a),
where lines with different $M$ deviate from each other.
Since $E_i^{\rm kin}(k) \propto \bar\epsilon^{2/3} k^{-5/3}$ is
predicted by the theory of Kolmogorov turbulence, we plot
$E_i^{\rm kin}(k) / \bar\epsilon^{2/3}$ in Fig.~\ref{f:ek}(b), where
the lines still deviate from each other, particularly around the
``bump'' ($k \sim 1$).
We next rescale the length by the domain size $\bar D$, which is shown
in Fig.~\ref{f:ek}(c).
Since $E_i^{\rm kin}(k) / \bar\epsilon^{2/3}$ has dimensions of
length to the power of $5 / 3$, it should be divided by
$\bar{D}^{5 / 3}$ in this rescaling, which corresponds to the shift of
the curves in Fig.~\ref{f:ek}(b) along the straight line $k^{-5/3}$.
We find from Fig.~\ref{f:ek}(c) that the rescaled plots seem to
collapse into a single universal curve, which follows a $-5/3$ power
law for $k \bar D \lesssim 1$ and has a ``bump'' at $k \bar D \gtrsim
2$.

The behavior in Fig.~\ref{f:ek}(c) can be interpreted as follows.
The scale of $k \bar D \lesssim 1$ corresponds to the inertial range,
in which the incompressible kinetic energy cascades into smaller
scales.
In the inertial range, the ``fluid'' consists of many domains, and the
movements of the domains produce vorticity distributions, as shown in
Figs.~2(b) and 2(e) in the main text, which cascade into smaller
scales.
This energy cascade stops at the scale of the domain size ($k \bar D
\sim 1$), since each domain itself does not contain quantized vortices
and the vorticity cannot cascade into a scale smaller than the domain
size.
Thus, the inertial range terminates at $k \bar D \sim 1$, and a
``bump'' emerges in the energy spectrum for larger $k \bar D$.
This situation is similar to the quantum turbulence in a
single-component superfluid, where a similar bump emerges at the scale
of the mean distance between quantized vortices~\cite{Nore}.
Below this scale, Kelvin-wave cascade occurs in the single-component
case.
In the present case, some excitation of domains may cause the energy
to cascade into smaller scales, which merits further study.
In Fig.~\ref{f:ek}(d), the energy spectra are further compensated by
$k^{-5/3}$, which indicates that the Kolmogorov constant is almost
unity.

Figure~\ref{f:ec} compares the kinetic energy spectra of the
compressible and incompressible velocity fields.
The compressible energy spectrum $E_c^{\rm kin}(k)$ is much smaller
than the incompressible energy spectrum $E_i^{\rm kin}(k)$ for all
scales, except the very small scale, which ensures that the
incompressible dynamics are dominant in the inertial range of
Kolmogorov turbulence.
\end{document}